\documentclass[12pt,a4paper]{article}
\usepackage{epsfig}
\def\eq#1{{Eq.~(\ref{#1})}}
\def\fig#1{{Fig.~\ref{#1}}}
\def\re#1{{Ref.~\cite{#1}}}

\newcommand{\beq}{\begin{equation}}
\newcommand{\eeq}{\end{equation}}
\newcommand{\di}[1]{\frac{\partial}{\partial #1}}
\newcommand{\beqar}[1]{\begin{eqnarray}\label{#1}}
\newcommand{\eeqar}{\end{eqnarray}}
\newcommand{\bas}{\bar{\alpha}_s}
\newcommand{\tN}{\tilde{N}}
\def\npb#1#2#3{    {\it Nucl. Phys. }{\bf B#1} (#2) #3}
\def\plb#1#2#3{    {\it Phys. Lett. }{\bf B#1} (#2) #3}
\def\prd#1#2#3{    {\it Phys. Rev. }{\bf D#1} (#2) #3}
\def\prep#1#2#3{   {\it Phys. Rep. }{\bf #1} (#2) #3}

\def\zpc#1#2#3{    {\it Z. Phys. }{\bf C#1} (#2) #3}

\def\sjnp#1#2#3{   {\it Sov. J. Nucl. Phys. }{\bf #1} (#2) #3}

\def\jetpl#1#2#3{  {\it JETP Lett. }{\bf #1} (#2) #3}
\def\epjc#1#2#3{   {\it Eur. Phys. J.}{\bf C#1} (#2) #3}

\def\ppsjnp#1#2#3{ {\it Sov. J. Nucl. Phys. }{\bf #1} (#2) #3}
\def\ppjetp#1#2#3{ {\it Sov. Phys. JETP }{\bf #1} (#2) #3}

\def\epjc#1#2#3{   {\it Eur. Phys. J.}{\bf C#1} (#2) #3}
\begin{document}
\title {{\bf New Scaling in High Energy DIS. }}
\author{
{\bf E.~Levin\thanks{e-mail: leving@post.tau.ac.il}}\quad {\bf and} \quad
{\bf K.~Tuchin\thanks{e-mail: tuchin@post.tau.ac.il}}
\\[10mm]
{\it\normalsize HEP Department}\\
{\it\normalsize School of Physics and Astronomy,}\\
{\it\normalsize Raymond and Beverly Sackler Faculty of Exact Science,}\\
{\it\normalsize Tel-Aviv University, Ramat Aviv, 69978, Israel}\\[0.5cm]
}

\date{December, 2000}
\maketitle
\thispagestyle{empty}

\begin{abstract}
We develop a new approach for solving the non-linear evolution
equation in the low $x_B$ region and show that the remarkable
``geometric" scaling of its solution
holds not only  in the saturation region, but in much wider 
kinematical region. This is in a full agreement
with experimental data (Golec-Biernat, Kwiecinski and Stasto).
\end{abstract}
\thispagestyle{empty}
\begin{flushright}
\vspace{-13.5cm}
TAUP-2659-2000\\
\today
\end{flushright}

\newpage


\section{Introduction.}

We believe that unitarity holds for any physical process. At very high
energies it manifests itself as a suppression of growth of cross sections  
as a function of energy. While at moderate energies
the linear
evolution equations hold, at higher energy corrections to those equations
arise which have essentially non-linear form.
It was suggested in \cite{GLR,LR,MIQU,MV} that there exists a certain
scale, called the saturation
scale $Q_S^2(x_B)$, at which those non-linear corrections set in.
This scale characterizes the high density phase of QCD which is
non-perturbative despite the smallness of the QCD coupling constant.
Consider the total cross section for deeply inelastic scattering of
virtual photon off the target $\sigma(Q^2,x_B)$. In the kinematical region
$Q^2\gg Q^2_s(x_B)>\Lambda^2$,where $\Lambda$ is a non-perturbative scale, 
the DGLAP evolution equations describe the
experimental data very well. In the infinite momentum frame, photon
interacts with only one parton in the partonic cascade.
 On the other hand, we expect that in the kinematical region
$Q^2<Q^2_s(x_B)$
(high density region) the virtual photon will most probably
interact simultaneously with at least two partons. Equation which
takes into account the possible simultaneous interactions of the
photon  with two partons  was derived in \cite{GLR,MIQU}
 in double logarithmic approximation and is written in parton language.
More than decade later it was shown by Balitsky\cite{BALITSKY}, that such
quadratic interactions describe the parton evolution in the whole
kinematical region (in the leading $\ln 1/x_B$) including very low $x_B$.
Recently, this result was independently  derived by Kovchegov\cite{KOV99}
in the framework of the dipole model\cite{DIPOLE,DIPOLE1}, by several
authors using
semi-classical approach\cite{CLASSIC,CLASSIC+}
and by Braun, using the standard form of the Pomeron-target
coupling\cite{BRAUN}. In
this paper we will consider the
non-linear evolution equation written in the dipole model picture. 

In the dipole model the deep inelastic scattering
of virtual photon off the
target has two consequent, well separated in time, stages: decay of the
photon into the system of
colour dipoles described by the wave function $\Phi(\vec x,z)$
and interaction of those dipoles with target with
amplitude $N(\vec x,y;\vec b_t)$,
where $\vec x$ stands for dipole of size $x$,
$b_t$ is an impact parameter and $y=\ln(x_B^0/x_B)$ is the rapidity
defined such that evolution starts at $y=0$.
We will assume that the typical transverse extent of the dipole amplitude
is much smaller then the size of the target and the typical impact
parameter $x\ll R,b_t$. Then
one can significantly simplify the impact parameter dependence of the
amplitude\cite{KOV00}. The high parton density evolution equation in the 
dipole approach now reads\cite{KOV99}
\begin{eqnarray}
\frac{\partial N(\vec{x}_{01}^2,y;\vec b_t)}{\partial y}&=&
-\frac{2\alpha_s N_c}{2\pi}\ln(\frac{\vec{x}^2_{01}}{\rho^2})
N(\vec{x}_{01}^2,y;\vec b_t)\nonumber\\
&+&\frac{\alpha_s N_c}{2\pi}\int_\rho d^2\vec{x}_2  \;
\frac{\vec{x}_{01}^2}{\vec{x}_{02}^2\vec{x}_{12}^2}\;
\left[2N(\vec{x}_{02}^2,y;\vec b_t)-N(\vec{x}_{02}^2,y;\vec b_t)
N(\vec{x}_{12}^2,y;b_t)\right]\quad,
\label{MAIN1}
\end{eqnarray}
where $\rho$ is ultraviolet cut-off. The kernel of this equation 
$$
|\Psi(\vec{x}_{01}\rightarrow\vec{x}_{02}+\vec{x}_{12})|^2=
\frac{x_{01}^2}{x_{02}^2 x_{12}^2}\quad,
$$
describes decay of the dipole $\vec{x}_{01}$ into two dipoles 
$\vec{x}_{02}$ and $\vec{x}_{12}$.
In the limit $N\ll 1$ this equation reduces to the BFKL one\cite{BFKL}.
The initial condition is taken to be of the Glauber
form\cite{GLAUB,KOV99}.
Once $N(\vec{x},y;\vec{b}_t)$ is known one can calculate the structure
function
as a convolution of it with the squared photon's wave function 
\cite{DIPOLE,DIPOLE1,NN,DIX}
\beq\label{F2S}
F_2(x_B,Q^2)=\frac{Q^2}{4\pi^2\alpha_{EM}}\int\frac{d^2\vec{x}}{2\pi}
dz\;
|\Phi(\vec{x},z)|^2\; \int\, d^2\vec{b_t}\; N(\vec{x},y;\vec{b_t})\quad ,
\eeq
where $z$ is a fraction of the photon's energy taken off by a struck
parton.
\eq{MAIN1} can be written in the momentum space as well.
Define,
following\cite{KOV00}, the two-dimensional Fourier transform of the amplitude
$N(\vec x,y)$:
\begin{eqnarray}
N(\vec{x},y)&=&\vec{x}^2\int_0^\infty dk\, k 
J_0(k x)\tN(k,y)\label{TILDAN}\\
\tN(k,y)&=&\int_0^\infty \frac{dx}{x}J_0(kx)N(x^2,y)\quad,
\end{eqnarray}
where the fact that neither \eq{MAIN1} nor initial condition (see
Ref.~\cite{KOV00}) depend
on a dipole direction was used to  integrate over polar angle
explicitly. Then \eq{MAIN1} can be written as
\beq\label{MAIN2}
\frac{\partial \tN(k,y)}{\partial y}= 
\bas\hat\chi\left(\hat\gamma(k)\right)\tN(k,y)
-\bas\tN^2(k,y)\quad,
\eeq
where $\hat\chi\left(\hat\gamma(k)\right)$ is an
operator such that
\beq\label{GAMMA}
\hat\gamma(k)=1+\frac{\partial}{\partial\ln k^2}
\eeq
is an operator corresponding to the anomalous dimension of the gluon
structure function
and the operator $\hat\chi$ corresponds to the following function 
\beq\label{CHI}
\chi\left(\gamma\right)=2\psi\left(1\right)-
\psi\left(1-\gamma\right)
-\psi\left(\gamma\right)
\eeq
which is an eigenvalue of the the BFKL equation.\footnote{Note,
that $\chi_{dipole}$ which was used by A. Mueller in the dipole
model and by Yu. Kovchegov in \re{KOV99} is different from 
that defined originally in BFKL papers \cite{BFKL}. The relation between
them is follows:
$2\chi_{dipole}(\lambda=2(1-\gamma))=\chi_{BFKL}(\gamma)
\equiv\chi(\gamma)$.
 $\lambda$
corresponds to the operator $\hat\lambda=-\frac{\partial}{\partial\ln k}$.}
We used convenient notation $\bas=\alpha_sN_c/\pi$.

It was pointed out by many authors\cite{MV,BL,LT}
 that in the high density region
(defined above) one expects that inclusive
observables will show remarkable scaling behaviour, which means that they
become a function of only one variable $Q^2/Q_S^2(x_B,b_t)$. It was shown
that both GLR and \eq{MAIN1} has this property\cite{BL,LT}.
In particular, we found
in \re{LT} the  scaling solution of \eq{MAIN1} in the
saturation region $Q^2\ll Q^2_S(x_B,b_t)$.   However, the
experimental
verification of this statement is quite difficult for technical reasons.
So, it was a great surprise when it turned out that this scaling behaviour
(so called ``geometric" scaling)
holds with 10\% accuracy in the \emph{whole} kinematical region
$x_B<0.01$\cite{GEOM}.

The goal of our paper is to show that indeed, the solution of the
\eq{MAIN2} scales with good accuracy in a wide high energy region.
We will begin by assuming {\it a priori} that such scaling solution exists.
In Sec.~2 we reduce \eq{MAIN2} to the non-linear one-dimensional equation by
introducing the scaling variable $\xi$. We
then suggest a model for the kernel $\chi$ of \eq{MAIN2} in the saturation
and diffusion kinematical regions.
In Sec.~3 we solve the one-dimensional (i.e.\  scaling) equation in the
framework of this model.
Then, in Sec.~4 we consider scaling-violating corrections to the scaling
solution, estimate numerically the size of those
corrections and found that they are small in the experimentally
accepted high energy kinematical region. Conclusions and discussion are
presented in Sec.~5.

\section{Definition of the problem.}

To proceed we have to specify the critical line $k^2=k^2_S(x_B)$ at which
shadowing corrections set in. In Ref.~\cite{LT} we found the critical
line by matching the double logarithmic solution of \eq{MAIN1} from the
kinematical region $\ln k^2\gg\alpha_sy\sim1$ (to the right of the
critical line in $(\ln k^2,y)$ coordinates) with saturating solution from
the region $\alpha_sy\gg\ln k^2$ (to the left of the critical line). It
reads:
\beq\label{CRL}
4\bas y=\ln\frac{k^2}{\Lambda^2}+\beta(b_t,A) \quad,
\eeq
where 
\beq\label{BETA}
\beta(b_t,A)=-2\ln S(b_t,A)-\frac{2}{3}\ln A\quad,
\eeq
$S(b_t,A)$ is target profile function  and $A$ is a number of nucleons in
the target.

Strictly speaking, \eq{CRL} is valid at sufficiently large values of $y$.
Indeed, the Glauber initial condition implies $Q^2_S(y=0)\sim
A^{1/3}$ while at large energies 
$Q^2_S(y)\sim\exp(4\bas y)A^{2/3}$. Hence, 
\eq{CRL} holds for $Q^2_S(y)\gg Q^2_S(y=0)$. 
Throughout this paper we assume that this condition is
satisfied\footnote{We are going to  discuss the $A$ dependence of the 
critical line at not too large $y$ in a separate publication.}. 

In the case of DIS on proton the good approximation for $S(b_t)$ is   
the  Gaussian profile function
\beq
S(b_t)=e^{-\frac{b_t^2}{R_P^2}}\quad.
\eeq
For  nuclear target the Woods-Saxon\cite{WS} profile function which can be
modeled by
\beq
S(b_t,A)=\theta(R_A-b_t)+\theta(b_t-R_A)\, e^{-\frac{b_t}{h}} 
\eeq
is usually used.
	  
Consequently, let us define the scaling variable
\beq\label{SCALVAR}
\xi =4\bas y-\ln\frac{k^2}{\Lambda^2}-\beta(b_t,A)  \quad.
\eeq
It was  shown in \re{BL} that as one approaches the saturation region,
scattering amplitude becomes a function of only one variable $\xi$.  
Since the scaling variable $\xi$ is defined up to some additive constant
we require that at $\xi\ge 0$ the amplitude  be a function
of only this variable. Hence, at $\xi<0$ one has to take into account
scaling-violating corrections which grow as $\xi$ gets smaller and
finally, at some small $\xi$ become of the same order as the
scaling solution, thus  destroying  the scaling behaviour. 
So, we look for the solution to the \eq{MAIN2} in the following form:
\beq\label{W1}
\tN(k,y;b_t)=\tN(\xi(k,y,b_t))+\delta\tN(k,y;b_t)\quad ,
\eeq
assuming that scaling-violating correction $\delta\tN(k,y;b_t)$  is small
perturbation of the scaling solution $\tN(\xi)$ at $\xi<0$ and vanishes at
$\xi\ge 0$. The boundary condition for the correction is  
\beq\label{BOUND}
\delta \tN(\xi=0,y;b_t)=\delta(y)\quad.
\eeq

It was argued in \re{BL} that anomalous dimension of the amplitude
equals $\gamma=\frac{1}{2}$ on the boundary of the kinematical
region where the amplitude is a function of
only one variable (this boundary is defined as $\xi$=0). This observation
provides an initial condition for the scaling
solution
\beq\label{INIT}
\frac{d\ln\tN(\xi)}{d\xi}\Big|_{\xi=0}=\frac{1}{2}\quad.
\eeq 

It is convenient to change variables in
\eq{MAIN2} $(y,k)\rightarrow(y,\xi)$ which means
the following substitutions
\beq
\frac{\partial}{\partial y}\rightarrow \frac{\partial}{\partial y}+
4\bas\frac{\partial}{\partial \xi}\quad;\qquad
\frac{\partial}{\partial \ln k^2}\rightarrow -\frac{\partial}{\partial
\xi} \quad.
\eeq
Using these formulae  one casts  \eq{MAIN2} to the form
\beq\label{GLAV}
\di y\tN(\xi,y;b_t)+4\bas\di\xi \tN(\xi,y;b_t)=
\bas\chi\left(1-\di\xi\right)\tN(\xi,y;b_t)
-\bas\tN(\xi,y;b_t)^2
\eeq

\begin{figure}
\begin{center}
\epsfig{file=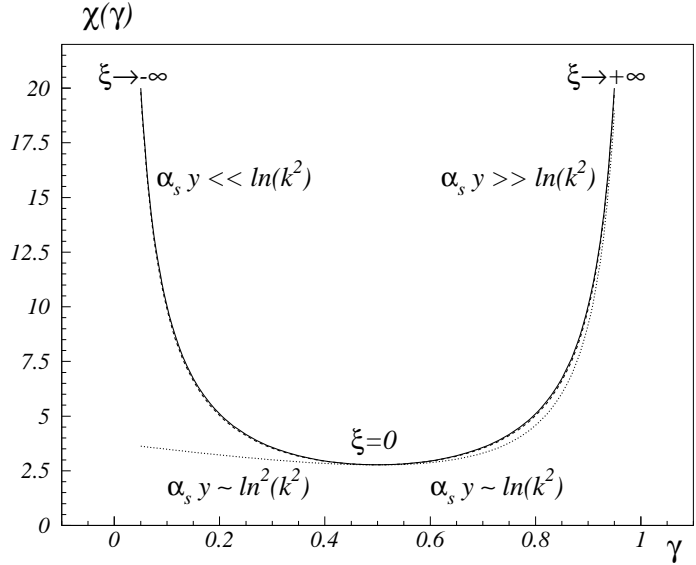,width=10cm,height=8cm}
\end{center}
\caption{{\it The eigenvalue of the BFKL evolution equation kernel as a
function of the anomalous dimension of the gluon structure function.
Solid line is the exact $\chi$ as given by \eq{CHI}, dashed line
corresponds to the model \eq{I2} and perfectly fits $\chi$ (it is almost
indistinguishable from solid line), the dotted
line is model for the right branch \eq{MODEL}.  The
different kinematical regions are shown.}}\label{FIG.CHI}
\end{figure}

\subsection{The model for the kernel.}
We do not know the exact analytical solution to the \eq{GLAV} even if we 
assume that $\tN$ is a function of only one scaling variable. To simplify
this equation we suggest the model for the function $\chi(\gamma)$. 
Note the following properties of this function which follows from
its definition \eq{CHI} and definition of the di-gamma function
$\psi(\gamma)$:
\begin{enumerate}
\item $\chi(\gamma)$ is defined in the region $0<\gamma<1$ 
(see \fig{FIG.CHI}). 
 \begin{itemize}
  \item $\gamma\rightarrow 0$
  corresponds to the double logarithmic approximation to the BFKL (or
  DGLAP) equation, i.e. $\ln k^2\gg\alpha_sy\sim 1$.
  \item $\gamma\rightarrow 1$ corresponds to the saturation region, i.e. 
  $\ln k^2\ll \alpha_s y\sim 1$.
  \item $\gamma\approx \frac{1}{2}$ corresponds to the diffusion
  approximation, i.e. $\ln^2 k^2\sim\alpha_sy\sim 1$.    
 \end{itemize}
\item \beq\chi(\gamma)=\chi(1-\gamma)\eeq.
\item
\beq\label{I1}
\chi(\gamma)=\frac{1}{\gamma}+2\sum_{n=1}^\infty
\zeta(2n+1)(1-\gamma)^{2n}
\eeq 
\item $\chi$ has minimum at $\gamma=\frac{1}{2}$,
$\chi(\frac{1}{2})=4\ln2$.
\end{enumerate}
Our model for $\chi$ in the whole region $1<\gamma<0$ is 
\beq\label{I2}
\chi(\gamma)=\frac{1}{\gamma}+\frac{1}{1-\gamma}+4\ln 2-4\quad .
\eeq
It is easily seen that this function satisfies properties 2 and 4. It 
has also correct asymptotic behaviour at the end points $\gamma\rightarrow
0,1$. Two last terms in the r.h.s. of \eq{I2} is an even polynom which
replaces the even
polynom in the r.h.s. of \eq{I1}.

In the scaling region  $\frac{1}{2}\le\gamma<1$ one can
expand $\gamma^{-1}$ term near some
point from $[0,1)$. This gives the term $\sim\gamma$. The model for
the right branch of $\chi$, that
satisfies property 2 and has correct asymptotic at
$\gamma\rightarrow 1$ reads    
\beq\label{MODEL}
\chi(\gamma)=\frac{1}{1-\gamma}+4\ln 2-4\gamma\quad.
\eeq
The main assumption of the model is that higher derivatives of the
amplitude are  much smaller than  the amplitude itself. In the next
section the scaling solution will be found which  justifies our
assumption. 

In the diffusion region $\xi\approx 0$ we
can expand $\chi$ near the point $\gamma\approx \frac{1}{2}$ 
\beq\label{DIFFUS}
\chi(\gamma)=4\ln 2+14\zeta(3)\left(\gamma-\frac{1}{2}\right)^2\quad .
\eeq
This is an approximation in which we  will calculate $\delta\tN(\xi,y)$
in Sec.~4.

\section{Solution to the scaling equation.}
Using \eq{W1} and \eq{MODEL} in \eq{GLAV} we get equation for the scaling
amplitude $\tN(\xi)$ at $\xi\ge 0$
\beq\label{SCALINGEQ}
\tN(\xi)-4(1-\ln2)\tN'(\xi)-2\tN(\xi)\tN'(\xi)=0\quad.
\eeq
\begin{figure}
\begin{center}
\begin{tabular}{cc} 
\epsfig{file=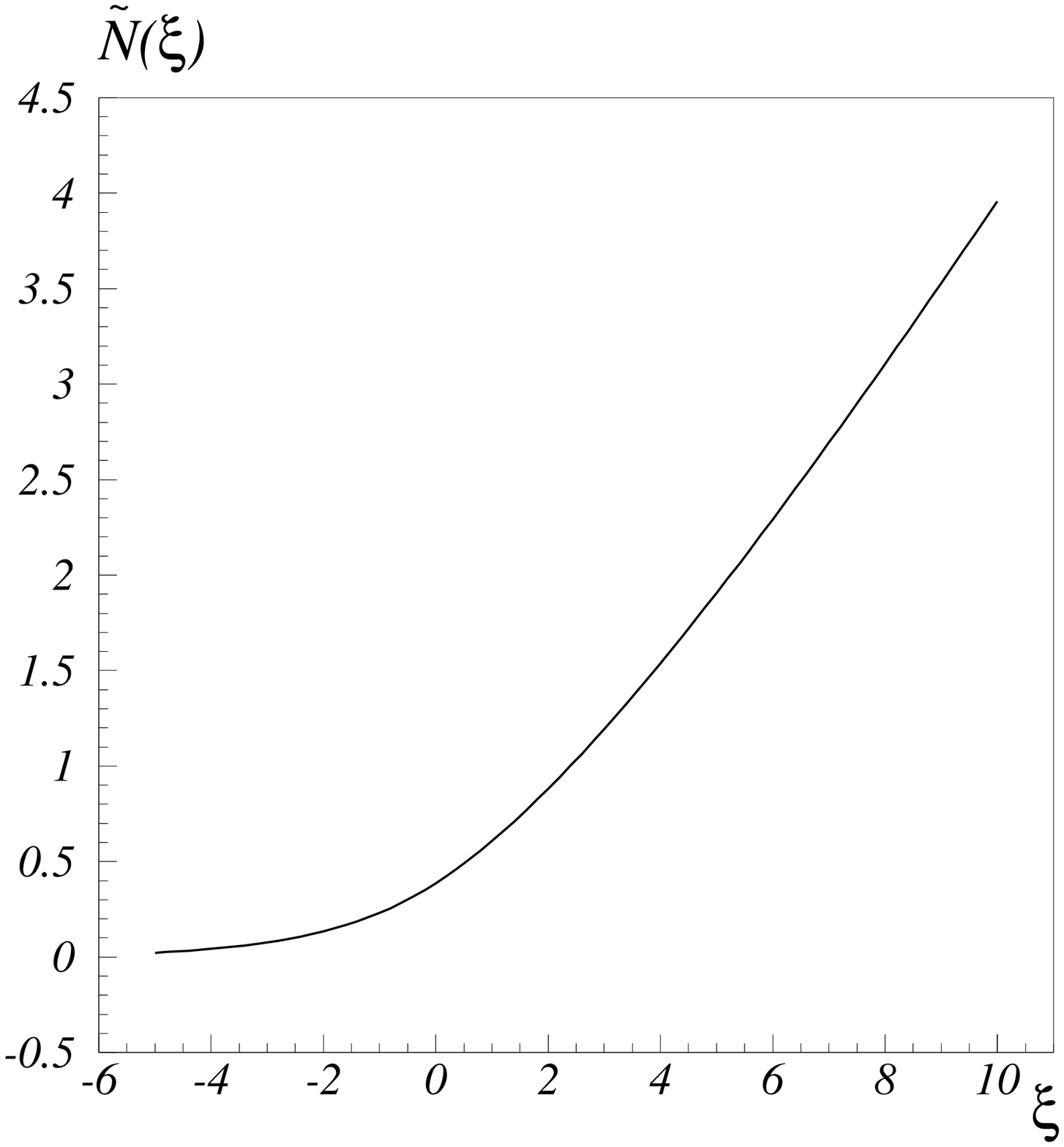,height=8cm,width=8cm}&
\epsfig{file=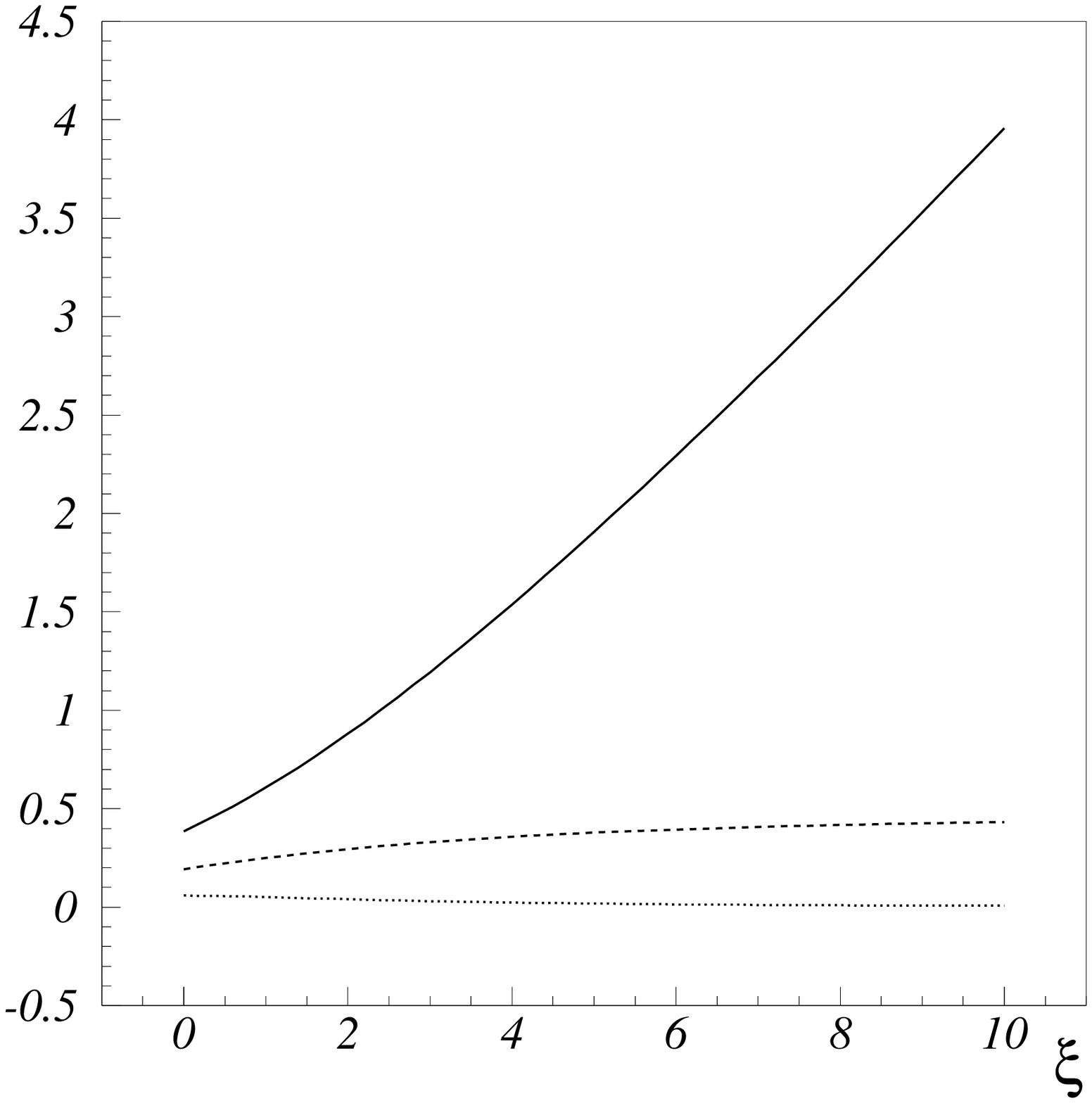,height=8cm,width=8cm}\\
(a)&(b)
\end{tabular}
\caption{{\it (a)~The scaling solution of the \eq{GLAV} in our model. At
$\xi>0$
this is a  numerical solution of \eq{SCALSOL} while at $\xi<0$ of
\eq{YYY}.
(b) The scaling solution (solid line), it's first (dashed line) and
second (dotted line) derivatives. 
}}\label{FIG.SCAL}
\end{center}
\end{figure}
Integration of  this equation yields
\beq\label{SCALSOL}
4(1-\ln2)\ln\tN+2\tN=\xi-\xi_0\quad .
\eeq
To find the value  of the integration constant $\xi_0$ we rewrite 
\eq{SCALINGEQ} in the form
\beq
\frac{d\ln\tN(\xi)}{d\xi}=\frac{1}{2\tN(\xi)+4(1-\ln2)}\quad.
\eeq
and use initial condition \eq{INIT} to get 
\beq\label{APP}
\tN(0)=2-4(1-\ln2)= 0.39\quad,
\eeq
and 
\beq
\xi_0=-2\tN(0)-4(1-\ln2)\ln\tN(0)=0.40\quad.
\eeq
By \eq{SCALSOL}, the asymptotic of the amplitude in the saturation region
is
\beq
\tN(\xi)=\frac{1}{2}\xi\quad.
\eeq

In the region $\xi<0$ we employ the diffusion approximation
\eq{DIFFUS}. The scaling equation then reads
\beq\label{YYY}
4\frac{d}{d\xi}\tN(\xi)=
4\ln2\tN(\xi)+14\zeta(3)\left(\frac{d}{d\xi}-\frac{1}{2}\right)^2
\tN(\xi)-\tN(\xi)^2\quad.
\eeq
Initial conditions for this equation are \eq{INIT} and obvious
requirement of the continuity $\tN(\xi\rightarrow-0)=
\tN(\xi\rightarrow+0)$. 

We show the numerical solution to the scaling equation in the whole
kinematical region in \fig{FIG.SCAL}~{(a)}. 
In \eq{MODEL} we neglected second and higher derivatives of the 
amplitude in comparison to  the first one and the amplitude itself.    
In the \fig{FIG.SCAL}~{(b)} it is shown that the neglection of the higher
derivatives was justified. Moreover, the statement that the forward
scattering amplitude is slowly varying function holds in general,
regardless of model, as scaling variable becomes positive and large.

\section{Corrections to the scaling solution.}

Now, as we know the scaling solution  of  \eq{MAIN2} at $\xi\ge0$, 
let us find
correction to this solution at $\xi<0$ due to the deviation from the
scaling. 
Substituting \eq{W1} into the \eq{GLAV}, employing diffusion approximation
\eq{DIFFUS} and keeping  terms
linear in perturbation $\delta\tN(\xi,y)$ we arrive at 
\begin{eqnarray}
-\frac{1}{\bas}\di y\delta\tN(\xi,y;b_t) &+&
14\zeta(3)\;\frac{\partial^2}{\partial\xi^2}\delta\tN(\xi,y;b_t)-
(4+14\zeta(3))\;\di\xi\delta\tN(\xi,y;b_t)\nonumber\\ 
&+&
(4\ln2+\frac{7}{2}\zeta(3)-2\tN(\xi))\;
\delta\tN(\xi,y;b_t)=0\quad,\label{CORREQ}
\end{eqnarray}
where $\tN(\xi)$ is the scaling solution at $\xi<0$. 
Let us define Melin transform $\delta\tN(\xi,\mu;b_t)$ of the
scaling-violating correction $\delta\tN(\xi,y;b_t)$ with
respect to the variable $\bas y$
\beq\label{MELIN}
\delta\tN(\xi,y;b_t)=\int_{a-i\infty}^{a+i\infty}\frac{d\mu}{2\pi i} \; 
e^{\mu\bas y}\delta\tN(\xi,\mu;b_t) \quad,
\eeq
where $a$ is situated to the right of all singularities of the integrand. 
Employing Melin transform one rewrites \eq{CORREQ} in the following form:
\begin{eqnarray}
14\zeta(3)\;\frac{\partial^2}{\partial\xi^2}\delta\tN(\xi,\mu;b_t)&-&
(4+14\zeta(3))\;\di\xi\delta\tN(\xi,\mu;b_t)\nonumber\\
&+&
(4\ln2+\frac{7}{2}\zeta(3)-2\tN(\xi)-\mu)\;\delta\tN(\xi,\mu,b_t)=0\quad.
\label{A10}
\end{eqnarray}

The boundary condition to this equation is specified by \eq{BOUND}.
We have to  show, however, that this boundary condition does
not contradict  the solution at $\xi>0$, i.e.\  the solution of 
the \eq{CORREQ} for the correction $\delta\tN(\xi,y)$ is
small in the  region $\xi\ge0$. Using \eq{MODEL} we get by analogy with
\eq{A10}
\beq
\di\xi\delta\tN(\xi,\mu;b_t)\left(\mu+4(1-\ln2)+2\tN(\xi)\right)
=\delta\tN(\xi,\mu;b_t)\left(1-2\tN'(\xi)\right)\quad.
\eeq
$\tN'(\xi)$ quickly approaches $\frac{1}{2}$ as $\xi$ increases, so,
indeed
neglection of $\delta\tN(\xi,y;b_t)$ at $\xi\ge0$ is justified.

Returning back to \eq{A10} we see, that non-linear term can be
neglected in the first approximation since $4\ln2+\frac{7}{2}\zeta(3)\gg
2\tN(\xi)$ (see \fig{FIG.SCAL}~{(a)}). Thus, we obtain the following
solution to \eq{CORREQ}:
\beq
\delta\tN(\xi,\mu,b_t)=e^{\left(\frac{1}{2}+\frac{1}{7\zeta(3)}\right)\xi}
\left(C_1(\mu,b_t)e^{\frac{\sqrt{\nu(\mu)}\xi}{7\zeta(3)}}+
C_2(\mu,b_t)e^{-\frac{\sqrt{\nu}\xi}{7\zeta(3)}}\right)\quad,
\eeq
where $C_1(\mu,b_t)$ and $C_2(\mu,b_t)$ have to be  chosen to satisfy the
boundary
condition \eq{BOUND}
\beq
C_1(\mu,b_t)=\bas \quad,\qquad C_2(\mu,b_t)=0\quad,
\eeq
and we introduced notation
\beq
\nu(\mu)=7\zeta(3)+1-14\zeta(3)\ln2+\frac{7}{2}\zeta(3)\mu\quad.
\eeq
Using \eq{MELIN} we obtain the final expression for the scaling-violating
correction
\beq
\delta\tN(\xi,y;b_t)=\frac{|\xi|}{\sqrt{\bas 8\pi 7\zeta(3)y^3}}
e^{\bas y(4\ln2-2-\frac{2}{7\zeta(3)})}
e^{\left(\frac{1}{2}+\frac{1}{7\zeta(3)}\right)\xi}
e^{-\frac{\xi^2}{8\bas 7\zeta(3) y}}\quad.
\eeq
To make the $b_t$ dependence of the correction manifest we rewrite it in
$(y,k^2)$ coordinates
\beq
\delta\tN(\xi,y;b_t)=\frac{|4\bas y-\ln k^2-\beta(b_t,A)|}
{\sqrt{\bas 8\pi 7\zeta(3)y^3}}
e^{4\ln2\bas y-\frac{1}{2}\ln k^2-\frac{\ln^2k^2}{56\zeta(3)\bas y}}
e^{-\frac{\beta}{2}\left(1+\frac{4\ln k^2}{56\zeta(3)\bas y}\right)
-\frac{\beta^2}{56\zeta(3)\bas y}} \quad.
\eeq
In the limit $\ln k^2\ll\alpha_s y$ it coincides with the solution to  the
BFKL equation in the diffusion approximation as it must be since we
neglected the non-linear term in \eq{CORREQ}.

\begin{figure}
\begin{center}
\begin{tabular}{c}
\epsfig{file=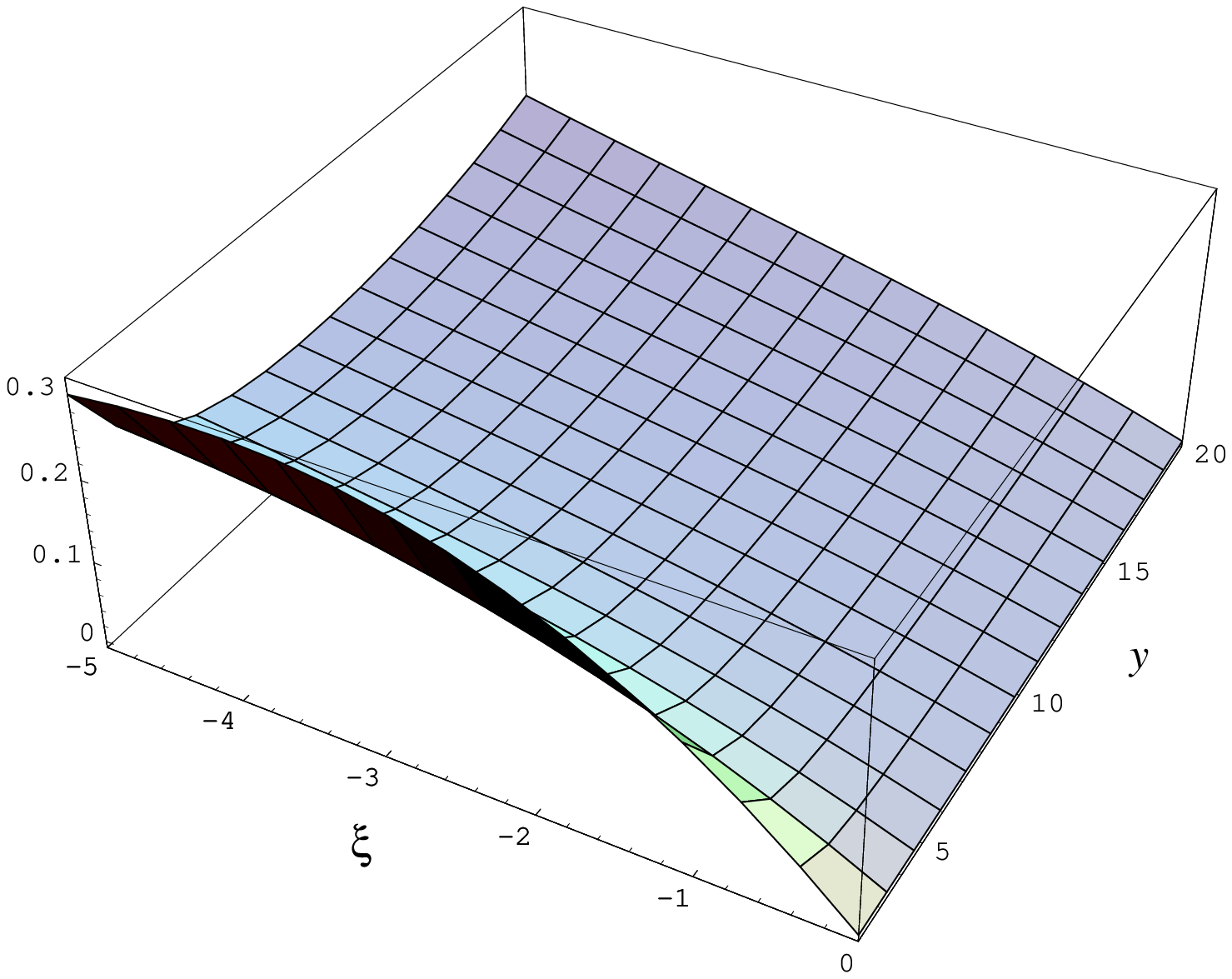, width=10cm,height=10cm}\\
\epsfig{file=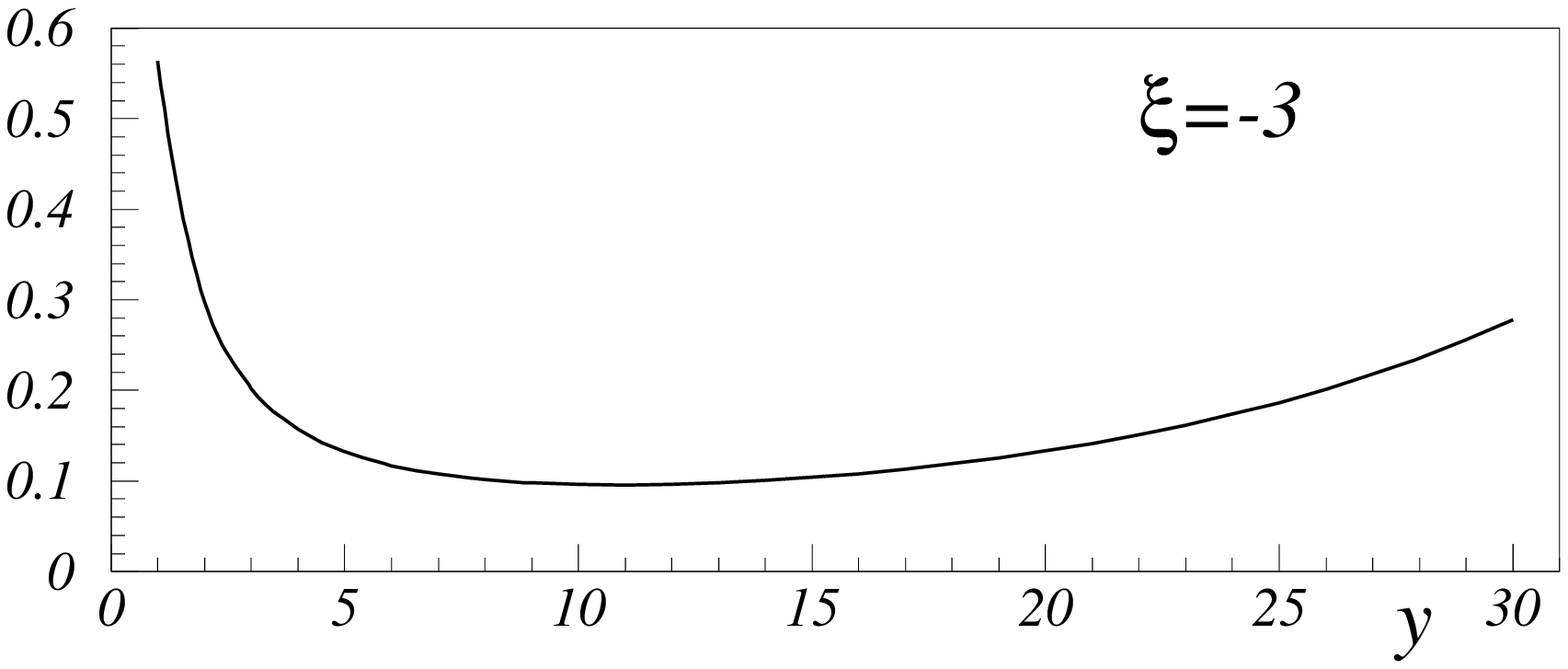,width=10cm,height=5cm}\\
\epsfig{file=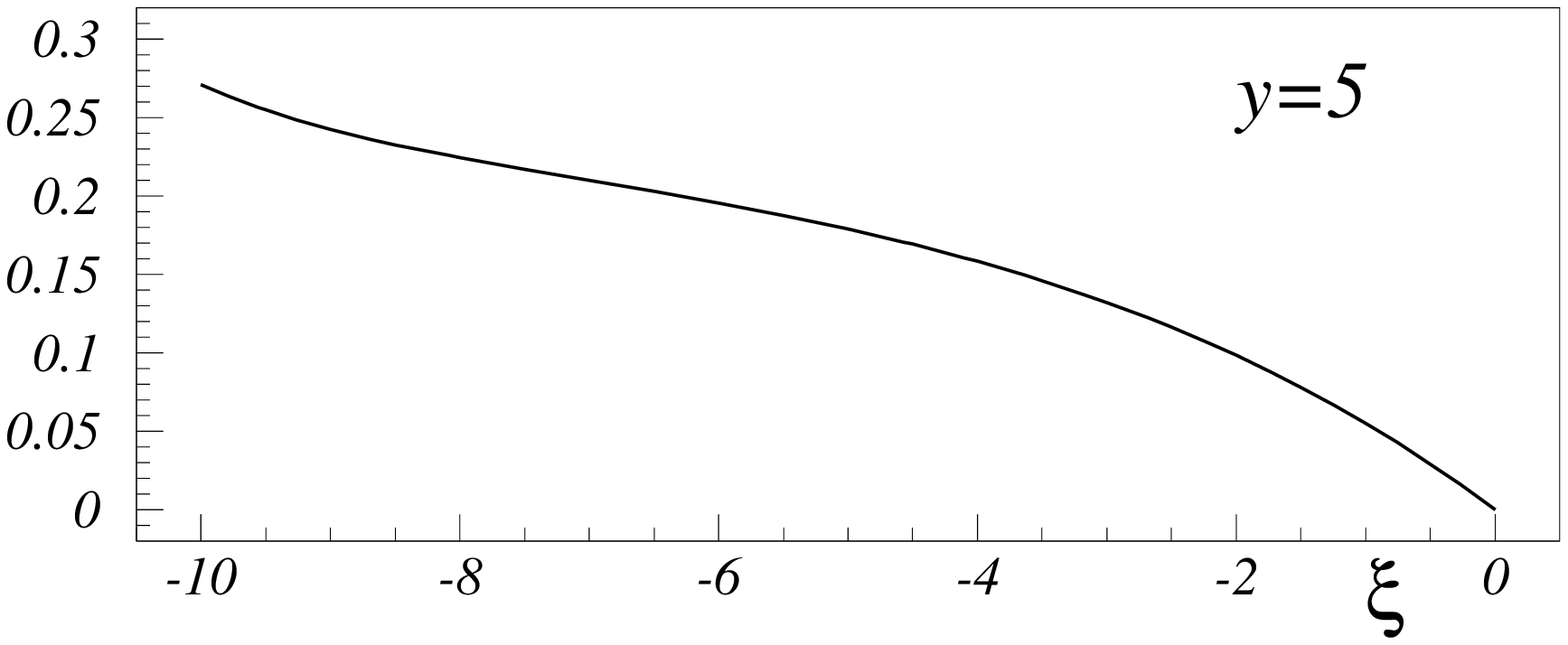,width=10cm,height=5cm}
\end{tabular}
\caption{{\it The ratio $\delta\tN(y,\xi)/\tN(\xi)$.}}\label{F3}
\end{center}
\end{figure}

The numerical value of the ratio $\delta\tN(y,\xi)/\tN(\xi)$ is
shown in \fig{F3}.
We see that there is a wide kinematical region where $\delta\tN(\xi,y)\ll
\tN(\xi)$. Correction increases in the following kinematical regions: 
$y\rightarrow\infty$ and $\xi\rightarrow-\infty$. Increasing of the
correction at $y\rightarrow 0$ is merely an artifact of the boundary
condition \eq{BOUND}.

\section{Discussion.}

\begin{figure}
\begin{center}
\begin{tabular}{cc}
\epsfig{file=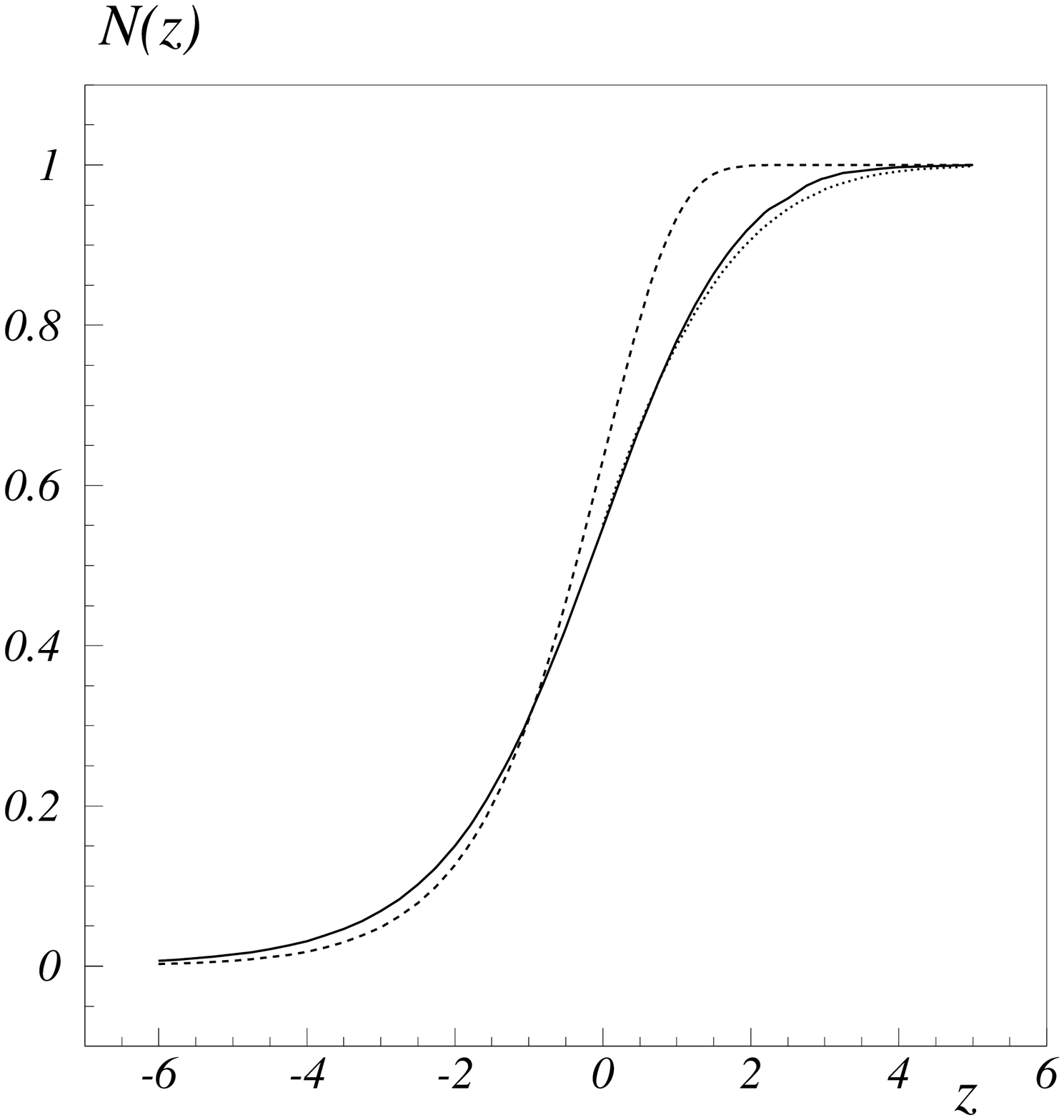, width=8cm,height=8cm}&
\epsfig{file=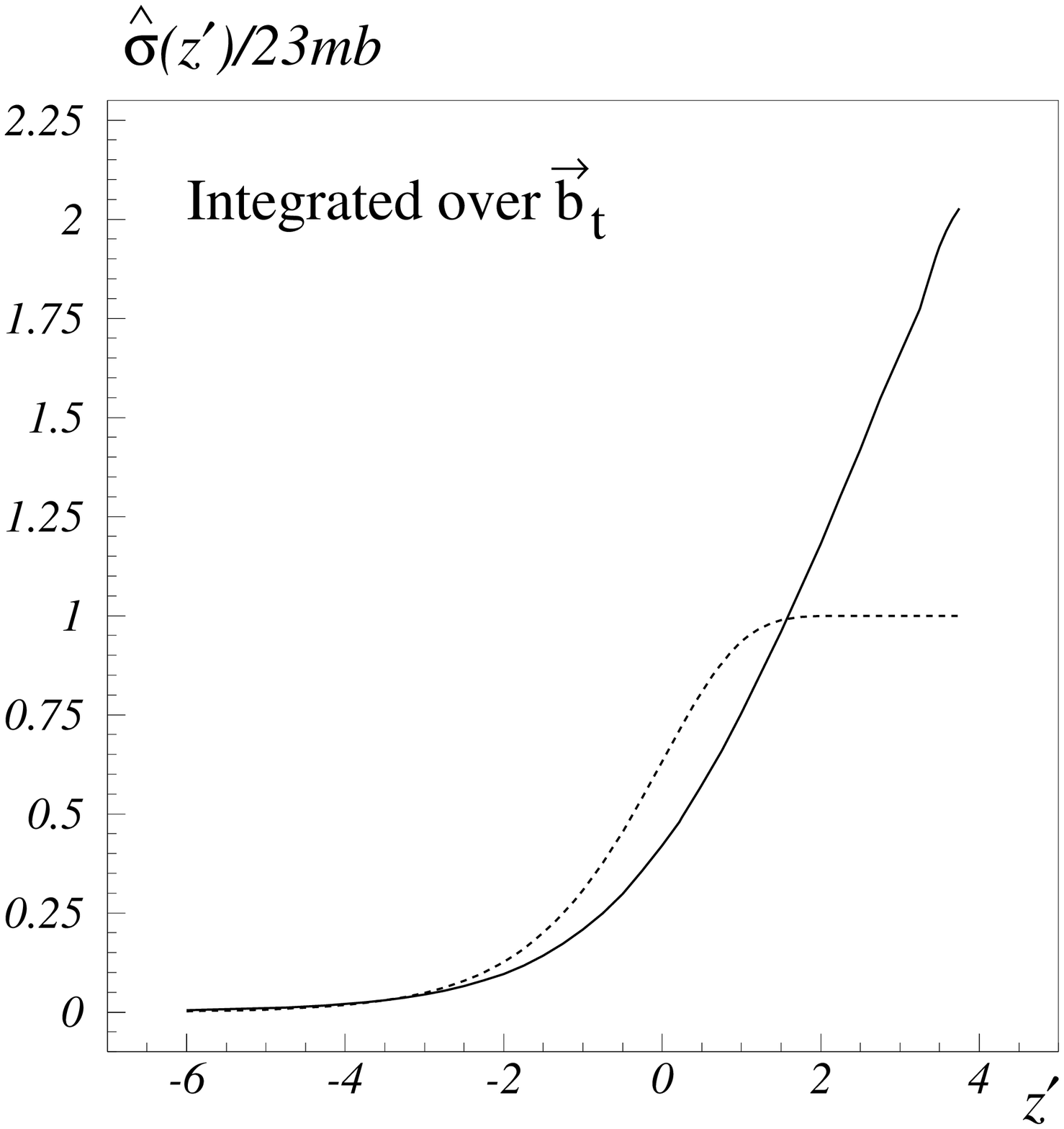, width=8cm,height=8cm}\\
(a)&(b)
\end{tabular}
\caption{{\it (a)Dipole -- target scattering amplitude $N(z)$ and 
(b) dipole -- target cross section $\hat\sigma(z')$ in the
scaling approximation versus scaling variable (a) $z$ and (b) $z'$: Solid
line is a Fourier
transform of $\tN(\xi)$ given in Fig.~2(a), dashed line is 
a Golec-Biernat -- Wusthoff model as explained in text and dotted line is
the $z\gg1$ asymptotic calculated in Ref.~\cite{LT}.}}\label{F4}
\end{center}
\end{figure}

In the previous two sections we have shown that in the wide kinematical
region $x_B<x_B^0$ the dipole -- target amplitude $\tN(k,y)$ is a
function of one variable $\xi$. For practical
uses we need to Fourier transform the amplitude to the
dipole-configuration space. Using \eq{TILDAN} one obtains
\beq
N(z)=e^z\int_0^\infty\;dt\, tJ_0(e^{z/2}t)\tN(-\ln t^2)\quad ,
\eeq
where we used \eq{SCALVAR}, then introduced a new integration variable
$t=\exp(-\xi/2)$ and defined
\beq
z=\ln\frac{\vec x^2}{\vec x^2_0}+4\bas y-\beta(b_t,A)
\eeq  
which is a dipole-configuration space scaling variable.
The result of numerical calculation is shown in \fig{F4} as well as the
result of our previous paper \cite{LT} where we found the asymptotic
$z\gg 1$ solution to the evolution equation. The dipole cross section is
introduced according to 
\beq
\hat\sigma(z')=2\int d^2\vec b_t N(z(\vec x,y,b_t))\quad, 
\eeq
where $z'=\ln\vec x^2/\vec x^2_0+4\bas y$.

Let us compare the results of our calculation with the successful
phenomenological model (for $A=1$) proposed by Golec-Biernat and
Wusthoff\cite{GBW}. 
Note, that they assumed that the $b_t$-dependence of the amplitude
factorizes out in the following way:
\beq
N(z)_{GBW}=N(z')_{GBW}\cdot\theta(b_{t0}^2-b_t^2)\quad,
\eeq  
where $2\pi b^2_{t0}=2\pi R_p^2=23$mb. We plotted $N(z)_{GBW}$ in the
\fig{F4}(a) and $\hat\sigma(z')_{GBW}$ in the \fig{F4}{b} (dashed curves).
It is seen that while the amplitudes in the \fig{F4}(a) are quite
close, the dipole cross section differ significantly as $z'$ becomes
positive and large. This difference is originated in $b_t$ integration 
and reflects the fact that our typical $b_t^2$ is of the order of 
$R^2\ln(Q_S^2(x_B)/Q^2)$\cite{LT}. 
Since in current experiments  the shadowing
corrections are still small, the closeness of the dashed and solid curves
in \fig{F4}(b) at $z'<0$ explains why such over-simplified model managed
to describe the experimental data well. However, in future experiments 
we will enter the region of $z>0$ where the model of Golec-Biernat and
Wusthoff does not work.

We understand the nature of the ``geometric" scaling phenomenon noticed
in \cite{GEOM}. While in the kinematical region of large $y$ and small 
$k^2$ (i.e.\ $z<0$) this scaling solution is an exact solution of the
evolution equation, at $z>0$ the scaling holds approximetely. We see in
\fig{F3} that corrections to the scaling behaviour are small in wide
kinematical region.

In conclusion, we would like to emphasize that our approach is developed 
for rapidities much larger then $y_A\sim\ln A^{1/3}$. 
In the forthcoming publication we are going to consider the ``geometric" 
scaling in DIS on heavy nuclei including rapidities $y\sim y_A$.


\vskip1.0cm
{\large\bf Acknowledgments}
\vskip0.3cm
We wish to acknowledge the interesting and useful disscusions with
S. Bondarenko, K. Golec-Biernat, E. Gotsman, Yu. Kovchegov, U. Maor and
L. McLerran.
This research was supported in part by the Israel Academy of Science and
Humanities and by BSF grant \#98000276.   


\newpage

\end{document}